\documentstyle[twoside,fleqn,espcrc2]{article}

\newcommand{\AmS}{{\protect\the\textfont2
  A\kern-.1667em\lower.5ex\hbox{M}\kern-.125emS}}

\title{On Big Bang Relics, the Neutrino Mass and the Spectrum of Cosmic Rays}

\author{Richard Wigmans\address{Department of Physics, 
        Texas Tech University, \\ 
        P.O. Box 1051, Lubbock TX 79409-1051, USA}}
       
\begin{document}
\def\NIM{Nucl. Instr. and Meth.~}
\def\etal{{\it et al.~}}
\def\eg{{\it e.g.,~}}
\def\ie{{\it i.e.~}}
\def\cf{{\it cf.~}}
\def\etc{{\it etc.~}}
\def\vs{{\it vs.~}}
\input epsf.sty

\begin{abstract}
It is shown that high-energy features of the cosmic ray spectrum, in
particular the kink around 4 PeV and the corresponding change in spectral index, may be
explained from interactions between highly energetic cosmic protons and relic Big Bang
antineutrinos, if the latter have a rest mass of about 0.4 eV/$c^2$. This explanation is 
supported by experimental data from extensive air-shower experiments, and in particular
by the observation (Fly's Eye) of a second kink around 300 PeV, and by the abrupt change
in the chemical composition of the cosmic ray spectrum that occurs at that energy. Both
facts follow naturally from our theory, which predicts additional verifiable features
of the cosmic ray spectrum in the few-PeV region, {\it e.g.} an abrupt decrease in
the $p/\alpha$ ratio.\end{abstract}

\maketitle

\section{Introduction}

In the 66 years that have passed since Pauli postulated the existence of neutrinos, a great deal has
been learned about these evasive particles. A large variety of experiments have been performed on neutrinos
produced  by nuclear reactors, by the Sun, by particle accelerators, in the Earth's atmosphere and in a
Supernova explosion. However, the most important (and in many respects the most interesting) source of
neutrinos remains unexplored until today: The Big Bang.   In my contribution to this conference, I will talk
about the relic Big Bang neutrinos, I will suggest a process through which they might reveal themselves, and I
will confront you with some experimental facts that seem to be strongly in support of this process.  

\section{Relic neutrinos}

According to the Big Bang model of the evolving Universe, large numbers of (electron) neutrinos
and antineutrinos have been around since the beginning of time. In the very first second,
when the temperature of the Universe exceeded 1 MeV, the density was so large that the
(anti-)neutrinos were in thermal equilibrium with the other particles that made up the
primordial soup: photons, electrons, positrons and nucleons. Photon-photon interactions
created $e^+e^-$ pairs, which annihilated into photon pairs. Interactions between
(anti-)neutrinos and nucleons turned protons into neutrons and vice versa.

This {\em leptonic era} came to an end when the density dropped to the point where the mean time between
subsequent $\nu$ interactions started to exceed the lifetime of the Universe, $\sim 1$ second after the Big
Bang. The production of neutrons through $\bar{\nu}_e p$ reactions had already stopped slightly earlier,
since it became energetically impossible. The (anti-)neutrinos {\em decoupled}
from the other particles in the Universe and the $n/p$ ratio determined the primordial
He/H ratio.

Since that moment, the wavelengths of the (anti-)neutrinos have been expanding in
proportion to the size of the Universe. Their present spectrum is believed
to be a momentum-redshifted relativistic Fermi--Dirac distribution:
\begin{equation}
N(p)~=~{1\over{\exp (pc/kT_{\nu}) + 1}}
\label{nuspec}
\end{equation}
characterized by a temperature $T_{\nu} = 1.95$ K, since 
$(T_\nu/T_\gamma)^3 = 4/11$ and $T_\gamma = 2.726 \pm 0.005$ K \cite{COBE}. 
The present density of these Big Bang relics is estimated at $\sim$ 100 cm$^{-3}$, for each neutrino flavor.
That is nine orders of magnitude larger than the density of baryons in the Universe. 

In spite of this enormous density, relic neutrinos have until now escaped direct
detection. The single most important reason for that is their
extremely small kinetic energy, which makes it difficult to find a process through which
they might reveal themselves. 

\section{How to detect relic neutrinos?}

Let us imagine a target made of relic $\bar{\nu}_e$s and let us bombard this target with
protons.
Let us suppose that we can tune this imagined proton beam to arbitrarily high energies,
orders of magnitude beyond the highest energies
that can be reached in our laboratories.
Then, at some point, the proton energy will exceed the value at which the center-of-mass
energy of the $p-\bar{\nu}_e$ system exceeds the combined mass of a neutron and a positron.
Beyond that energy, the inverse $\beta$-decay reaction
\begin{equation}
p + \bar{\nu}_e \rightarrow n + e^+
\label{ibeta}
\end{equation}
is energetically possible.

The threshold proton energy for this process depends on the mass of the $\bar{\nu}_e$
target particles. If this mass is large compared to the $10^{-4}$ eV kinetic energy typically
carried by the target particles, this may be considered a stationary-target problem, and
the center-of-mass energy of the $p-\bar{\nu}_e$ system can be written as
\begin{equation}
E_{\rm cm} = \sqrt{m_p^2 + m_\nu^2 + 2 E_p m_\nu} \approx \sqrt{m_p^2 + 2 E_p m_\nu} 
\label{com1}
\end{equation}
since $m_\nu \ll m_p$.  
When the experimental mass value of the proton (938.272 MeV) is substituted in
Equation \ref{com1} and the requirement is made that $E_{\rm cm} > m_n + m_e$ (940.077 MeV), this
leads to  
\begin{equation}
E_p m_\nu > 1.7 \cdot 10^{15}~ {\rm (eV)}^2 
\label{com2}
\end{equation}
This process will thus take place when
\begin{equation}
E_p ({\rm eV}) > {1.7 \cdot 10^{15}\over m_{\nu} ({\rm eV})} 
\label{com3}
\end{equation}

In our {\em Gedanken experiment}, this threshold would reveal itself through a decrease in the fraction
of beam protons that traversed the target without noticing its presence, as $E_p$ is
increased beyond the threshold. Is there any experimental evidence for a phenomenon of this type?

\section{The cosmic knee}
 
Obviously, the target needs to be huge in order to see any effect of the described
process. The largest target we can imagine is the entire Universe, which is indeed filled
with $\bar{\nu}_e$s. This target is continuously traversed by protons, and by other atomic
nuclei which together constitute the cosmic rays. 
\begin{figure}[htb]
\vspace{-0.7cm}
\epsfxsize=76mm \epsfbox{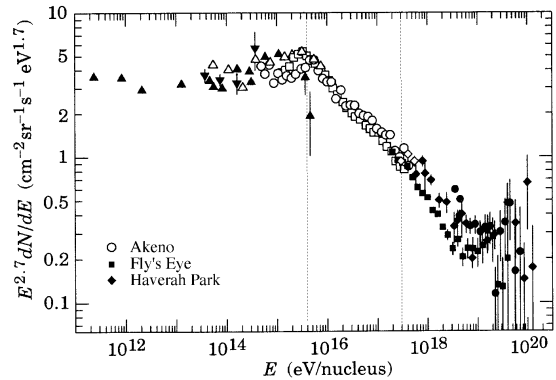}
\vspace{-1.0cm}
\caption{\small
The all-particle cosmic-ray spectrum}
\label{Kink1}
\end{figure}
\vskip -6mm         
Figure \ref{Kink1} shows the spectrum of these cosmic rays, compiled by the
Particle Data Group \cite{PDG2}. The spectrum falls extremely steeply with energy.
In general, it is well described by a power law
\begin{equation}
{dN\over dE}~\sim~ E^{-n}
\label{cosray}
\end{equation}
with $n \approx 2.7$ for energies below 1 PeV.
In order to display characteristic features of this spectrum which would
otherwise be hard to discern, the differential energy spectrum has been
multiplied by $E^{2.7}$ in Figure \ref{Kink1}. The steepening that occurs between 1 PeV and 10 PeV, where the
spectral index $n$ changes abruptly from 2.7 to 3.0, is known as the {\em knee} of the cosmic ray spectrum. 
Ever since its discovery, this feature has been the subject of intense interest. 

We notice that this knee exhibits exactly the features that we expected to see in our Gedanken
experiment: The particle flux suddenly starts to decrease when the threshold is passed.   
Therefore, we postulate the following hypothesis:

{\em The change of the spectral index in the all-particle cosmic ray spectrum at an energy of 
$\sim 4 \cdot 10^{15}$ eV is caused by the onset of the reaction $p + \bar{\nu}_e \rightarrow
n + e^+$, which becomes energetically possible at this point.}

This hypothesis necessarily implies (Equation \ref{com3}) that the mass of the electron neutrino equals 
0.4 eV. Also, the knee would have to be an {\em exclusive feature of the proton component} of the cosmic ray
spectrum, if the hypothesis were correct. Beyond 4 PeV, one would thus expect to see a gradual drop in, for
example, the $p/\alpha$ event ratio, and this could be experimentally verified.

\section{A second knee?!}

If protons interact with the relic background neutrinos, other cosmic ray particles may as well.   The
equivalent reactions in which $\alpha$ particles are dissociated in collisions with relic neutrinos and
antineutrinos  
\begin{eqnarray}
\alpha + \nu_e~ \rightarrow~ 3p + n + e^-\\
\alpha + \bar{\nu}_e~ \rightarrow~ p + 3n + e^+
\end{eqnarray}
have $Q$-values of 27.5 MeV and 30.1 MeV, respectively. The threshold
energies for these reactions are larger than the threshold energy
for reaction (\ref{ibeta}) by factors of 61 and 66, respectively.  
The cosmic ray spectrum in the range around the ``knee'' was measured in detail by the
Akeno extensive air-shower experiment\cite{akeno}. From a fit to their experimental data,
the kink was located at $\log{E_p}~({\rm eV}) = 15.67$.   
 
If the above explanation of the origin of the knee is correct, a kink in the
spectrum of cosmic $\alpha$ particles should thus be expected at an energy that is larger by
a factor of 61-66: $\log{E_\alpha} \approx 17.5$ ($E_\alpha \approx 3\cdot 10^{17}$ eV).  
It is remarkable that several extensive air-shower experiments that have
studied the cosmic ray spectrum at these high energies have reported a kink in the
area around $\log{E} = 17.5$. The Fly's Eye detector, which obtained the largest event
statistics, observed a change in the spectral index from $3.01 \pm 0.06$ for energies
$< 10^{17.6}$ eV to $3.27 \pm 0.02$ for energies $10^{17.6} < E < 10^{18.5}$ eV\cite{fleye}.
The Haverah Park detector also reported a kink at $\log{E} = 17.6$, with the spectral index
changing from $3.01 \pm 0.02$ to $3.24 \pm 0.07$\cite{haverah}. 
\begin{figure}[htb]
\epsfxsize=76mm \epsfbox{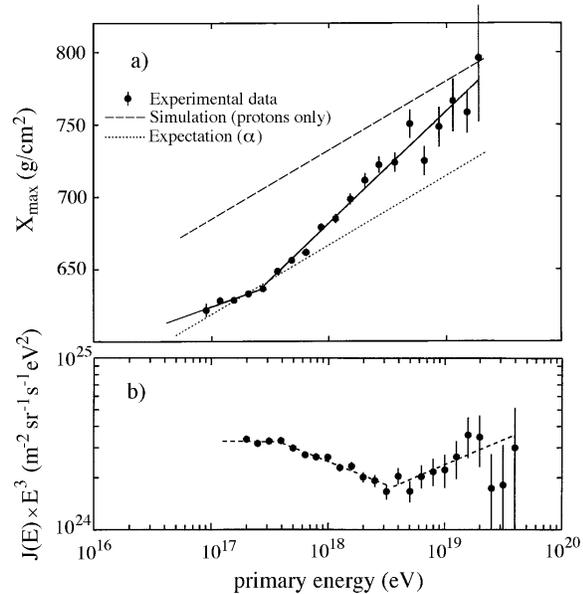}
\vspace{-0.8cm}
\caption{\small
Experimental results obtained with the Fly's Eye detector\protect\cite{fleye}. See text for
details.}
\vskip -7mm
\label{fly}
\end{figure}

The differential energy spectrum measured with the Fly's Eye detector is shown in Figure \ref{fly}$b$. 
The contents of the energy bins have been multiplied by $E^3$ in this figure (compared to $E^{2.7}$ in
Figure \ref{Kink1}). The abrupt change of the spectral index $n$ from 3.01 to 3.27 becomes very clear in
this way. 

The Fly's Eye detector is also capable of measuring the {\em altitude} at which the shower
maximum is located. Figure \ref{fly}$a$ shows the amount of material (air) penetrated by the
shower particles when they reach the shower maximum, $X_{\rm max}$, as a function of the shower
energy. It is remarkable that this distribution exhibits an unmistakable kink at approximately
the same energy as the energy spectrum itself. This indicates a clear change in the chemical
composition of the cosmic rays at that point. 

The dashed curve in Figure \ref{fly}$a$
represents the results of a simulation modeling showers initiated by protons\cite{fleye}.
The equivalent curve for showers induced by $\alpha$
particles can be derived from this in a straightforward manner. Two effects have to be taken into
account. Both effects cause $X_{\rm max}$ to be smaller for $\alpha$s than for protons of the same energy.

The first effect concerns
the {\sl interaction length} ($\lambda_{\rm int}$) which for $\alpha$s is shorter than
for protons. The interaction length of protons in air amounts to 90 g/cm$^2$.
Since the total cross section for interactions of highly energetic ions with atomic
number
$A$ scales with $A^{2/3}$, $\lambda_{\rm int}$ scales with $A^{-2/3}$. This gives
$\lambda_{\rm int} = 55$ g/cm$^2$ for $\alpha$s in air.

The second effect concerns the {\sl particle multiplicity}.
In the first nuclear collision, the energy carried by the incoming $\alpha$-particles is distributed to a
larger number of secondary particles. In the subsequent shower
development, the shower maximum is thus reached earlier than in case of showers induced by protons
of the same energy. The effect of this can, in first approximation, be estimated by
considering the ion-induced shower as a collection of $A$ separate showers, each with an energy
$E_A/A$. From the slope of the dashed line in Figure \ref{fly}$a$, one may estimate 
$\alpha$-induced showers in air to reach their maximum earlier than proton-induced ones, by 30
g/cm$^2$, counting from the first nuclear interaction.
 
These (somewhat oversimplified) considerations thus lead to $X_{\rm max}$ values for 
$\alpha$-induced reactions in the atmosphere that are 65 g/cm$^2$ smaller than the values for protons
of the same energies, as indicated by the dotted curve in Figure \ref{fly}$a$.    
\vskip 2mm

The experimental data shown in Figures \ref{fly}$a$ and \ref{fly}$b$ are in complete and detailed agreement
with the expectations following from our hypothesis.
These expectations can be summarized as follows:
\begin{itemize}
\item At energies below $4\cdot 10^{15}$ eV, the chemical composition of the cosmic rays is constant.
Protons are the most dominant component. The average shower maximum follows the dashed
line (Figure \ref{fly}$a$), as required by the well-known logarithmic energy dependence. The spectral index
$n = 2.7$.
\item At $4 \cdot 10^{15}$ eV, the threshold for reaction (\ref{ibeta}) is crossed. The spectral index
changes from 2.7 to 3.0. Beyond the threshold, protons gradually disappear from the spectrum, which becomes
more and more dominated by $\alpha$s. As $E$ increases, the shower maximum thus moves away from the
dashed line to the dotted line in the $X_{\rm max}$ plot.
\item  For energies around $10^{17}$ eV, $\alpha$s have replaced protons as the most dominant component. The
spectral index is still 3.0
\item At $3 \cdot 10^{17}$ eV, a new threshold is crossed, this time for reactions (8)
and (9). The spectral index changes from 3.0 to 3.25. 
Beyond the threshold, $\alpha$s gradually disappear from the spectrum. In that process,
protons carrying a significant fraction of the energy are produced. This leads to a gradual
enrichment in protons. Therefore, in the $X_{\rm max}$ plot, the position of the shower maximum moves back
from the dotted line to the dashed line, as the energy increases.
\item At $10^{18.5}$ eV, a new threshold is crossed,  known as the {\em ankle} of the cosmic-ray
spectrum.  At this energy, the gyromagnetic radius of a proton starts to exceed the galactic radius and
therefore extragalactic protons may reach the Earth. The spectral index changes back to its original value of
2.7. The chemical composition continues to change in the same sense as before, \ie, a further gradual
enrichment in protons. 
\end{itemize}
\begin{figure}[htb]
\vspace{-0.6cm}
\epsfxsize=76mm \epsfbox{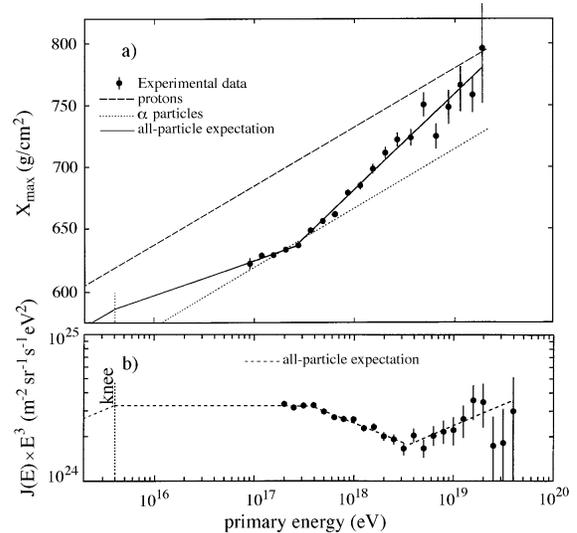}
\vspace{-0.8cm}
\caption{\small
The expected average depth of the shower maximum in the atmosphere ($a$) and the expected spectral index
($b$), as a function of the energy of the cosmic rays. Experimental data were obtained with the Fly's Eye
detector \protect\cite{fleye}. See text for details.}
\vskip -7mm
\label{fly2}
\end{figure}

This sequence of effects is graphically shown in Figure \ref{fly2}, which is an extension of Figure
\ref{fly} and includes the knee area. The experimental data are in excellent
agreement with these predictions. 

\section{How is this all possible?}

We now turn our attention to the million-dollar question: {\em How can the process that is the basis
of our hypothesis play such a significant role, given its extremely small cross section?} 

The cross section for the inverse process, $\bar{\nu}_e$ scattering off protons, was measured to be
proportional to the {\em available} energy in the center-of-mass system,
\ie the energy above the $Q$-value of 1.805 MeV\cite{Perkins}:
\begin{equation}
\sigma~=~0.34 (E_{\rm cm} - Q) 10^{-41} {\rm cm}^2
\label{xsec}
\end{equation}
with the energies expressed in units of MeV. Therefore, the cross section rises almost
proportional to the energy of the incoming protons in the energy range between $10^{16}$ eV
($\sigma \approx 10^{-41}$ cm$^{2}$) and $10^{18}$ eV ($\sigma \approx 10^{-39}$ cm$^2$).
For a target density of $\sim 100$ cm$^{-3}$, these cross sections translate in mean free paths of
$10^{37 - 39}$ cm, or average lifetimes of $10^{19 - 21}$ years, \ie\ 9 -- 11 orders of magnitude longer than
the age of the Universe.
If this is all there is, the high-energy cosmic ion spectra could thus never have been affected at a
significant level by the hypothesized process.
\vskip 3mm

However, it is important to realize that, with a mass of 0.4 eV, the relic
$\bar{\nu}_e$s would be {\em nonrelativistic} ($kT \sim 10^{-4}$ eV). 
Typical velocities would be of the order of $10^5$ m/s in that case\cite{TG}, less than the 
escape velocity from the surface of the Sun.
Such neutrinos may be expected to have accumulated in gravitational potential wells.
Weiler\cite{weiler} recently estimated that the density of relic neutrinos in our own galaxy
would increase by four orders of magnitude (compared to the universal density of 100 cm$^{-3}$)
if their mass was 1 eV.

Locally, this effect could be much more spectacular. Extremely dense objects, such as neutron
stars, black holes or active galactic nuclei (AGN) could accumulate very large numbers of relic neutrinos
and antineutrinos in their gravitational field. 
Let us consider, as an example, a typical neutron star, with a mass ($M$) of $3 \cdot 10^{30}$
kg and a radius of 10 km. Even at a distance ($r$) of one million kilometers from this object,
the escape velocity is still considerably larger than the typical velocity of these relic
neutrinos: 700 km/s.

The concentration of relic neutrinos in such a local potential well is ultimately
limited by the Pauli principle, which limits their phase-space density to $4 g_\nu h^{-3}$,
where
$g_\nu$ denotes the number of helicity states and $h$ Planck's constant\cite{TG}. 
Since the escape velocity scales with $r^{-1/2}$, the maximum neutrino density is proportional
to $r^{-3/2}$, and reaches values of $\cal{O}$$(10^{12}/{\rm cm}^3)$ near the surface of this
neutron star. If the source of the potential well has a larger mass, the
achievable neutrino densities increase proportional to $M^{3/2}$. In the ``neutrino
atmosphere'' surrounding an AGN with a mass of $10^6 M_\odot$, the lifetime of a $10^{16}$ eV proton  
could thus even be reduced to only a few years.
      
The example of a neutron star was chosen since objects of this type may well play a role in
accelerating protons to the very high energies considered here. Neutron stars
usually rotate very fast and exhibit very high magnetic fields (typically $\sim 10^8$ T).
When the magnetic axis does not correspond to the rotation axis, the changing magnetic
fields in the space surrounding the neutron star may give rise to substantial electric fields,
in which charged particles may be accelerated to high energies\cite{KOT91}.
The strong magnetic fields, in turn, will tend to confine the accelerated particles to the
region surrounding the neutron star. When the charged particles are electrons, the acquired
energy is emitted in the form of synchrotron radiation, a characteristic signature of many
pulsars.

However, for protons this type of energy loss only plays a
significant role at energies that are 13 orders of magnitude ($(\gamma_p/\gamma_e)^4$)
larger than for comparable electrons. For example, a 10$^{16}$ eV proton describing a circular
orbit with a radius of 1,000 km loses per orbit about 0.6\% of its energy in the form of
synchrotron radiation. To put this in perspective, that is about a factor of five less than
the fractional energy loss experienced by 100 GeV electrons in CERN's LEP ring.
 
It is, therefore, not inconceivable that a stationary situation might arise in which protons
and heavier ions, initially attracted by gravitational forces, and accelerated to high
energies in the electric fields mentioned above, end up in a closed orbit around
the dense object (neutron star, black hole, AGN) in which energy gained and synchrotron losses balance each
other. In that case, the ions could spend a long time in the dense neutrino atmosphere surrounding
the gravitational source and collisions with relic neutrinos might become a realistic possibility.
\vskip 2mm

The idea that an AGN might act as the engine that produces cosmic rays is not at all new.
If this engine works as a synchrotron, as described above,
there are two ways in which the trapped accelerated ions can escape to other regions of the galaxy: Through
collisions with other nuclei or through collisions with relic neutrinos. In the first case, the
reaction products acquire transverse momentum and are scattered out of the acceleration plane. However, if
a proton collides with a relic $\bar{\nu}_e$, the resulting neutron has approximately the same
four-vector as the original proton. The escaping neutron thus remains in the acceleration plane. That also
holds for the proton into which this neutron decays after a while (100 years or so!). If the Earth was not
located in the AGN's acceleration plane, these protons would thus never reach us, in contrast with the protons
scattered in nuclear collisions. The measured proton spectrum would thus exhibit a kink at the threshold for
the $p\bar{\nu}_e$ reaction.
  
\section{Conclusions}

The high-energy cosmic ray spectrum exhibits some intriguing features that can be explained in a
coherent manner from interactions between cosmic protons or $\alpha$ particles and relic $\bar{\nu}_e$s if
the latter have a restmass of 0.4 eV/$c^2$: 
\begin{itemize}
\item Kinks at $4\cdot 10^{15}$ eV and $3\cdot 10^{17}$ eV, which would correspond to the thresholds for the
$p\bar{\nu}_e$ and $\alpha \bar{\nu}_e$ reactions.
\item The energy separation between these kinks is consistent with the difference
between the $Q$-values of the $p\bar{\nu}_e$ and $\alpha \bar{\nu}_e$ reactions.
\item The change of the chemical composition at $3\cdot 10^{17}$ eV, where $\alpha$s start to disappear
from the spectrum and the protons reappear.
\end{itemize}

If collisions with relic neutrinos were indeed responsible for these features, a large concentration of
such neutrinos should be present in the vicinity of the source of the high-energy cosmic baryons.
The required conditions could be met if these baryons were accelerated in the electromagnetic fields
surrounding an AGN.

There are various ways to test this hypothesis experimentally. First, one could
explicitly measure the chemical composition of the cosmic-ray spectrum in the energy region from 10$^{14}$ to
$10^{18}$ eV. Second, one could measure the shower-maximum distributions for energies below the Fly's
Eye lower limit to see if the predicted kink near 4 PeV (Figure \ref{fly2}$a$) corresponds to reality.

It should be emphasized that the predictions derived from our hypothesis are completely at variance with
those derived from the ``Standard Model of Cosmic Ray Physics'' \cite{Blanford}. In this model, the cosmic
rays are the result of particle acceleration in the shock waves produced in supernova explosions. This model
predicts, for example, a maximum energy proportional to the nuclear charge $Z$ of the particles. In this
model, the ``knee'' would thus be a feature of the high-$Z$ component of the cosmic rays, while we predict
it to be a feature of protons and protons alone.   

Measurements of the high-energy cosmic proton and $\alpha$ spectra would thus not only constitute a
crucial test of our hypothesis, but also check some key predictions of this ``Standard Model''.
 
\bibliographystyle{unsrt}

\end{document}